\documentstyle[aps,amssymb,eqsecnum,epsf]{revtex}
\begin{document}



\title{Notes on causal differencing in ADM/CADM formulations: a 1D comparison}

\author{
Luis Lehner${}^{1}$, Mijan Huq${}^{2}$ and David Garrison${}^{2}$
}

\address{
${}^{1}$Center for Relativity,\\
The University of Texas at Austin, Austin, TX 78712 \\
${}^{2}$Department of Astronomy \& Astrophysics, and\\
Center for Gravitational Physics \& Geometry,\\
The Pennsylvania State University, University Park, PA 16802}

\date{\today}

\maketitle

\begin{abstract}
Causal differencing has shown to be one of the promising and successful 
approaches towards excising curvature singularities from numerical
simulations of black hole spacetimes. So far it has only been
actively implemented in the ADM and Einstein-Bianchi 3+1 formulations of the
Einstein equations.
Recently, an approach closely related to the ADM one, commonly referred to 
as as ``conformal ADM'' (CADM) has shown excellent results when modeling 
waves on flat spacetimes and black hole spacetimes where singularity avoiding
slices are used to deal with the singularity. In these cases, the use of CADM 
has yielded longer evolutions and better outer boundary dependence than those 
obtained with the ADM one. If this success translates to the case where 
excision is implemented, then the CADM formulation will likely be a prime 
candidate for modeling generic black hole spacetimes.  In the present work we 
investigate the applicability of causal differencing to CADM, presenting the 
equations in a convenient way for such a goal and compare its application with 
the ADM approach in 1D.
\end{abstract}

%
%
%


\section{Introduction}
\label{sec:Introduction}
One of the goals of numerical relativity that has proven to be elusive (using a 3+1 
splitting of the Einstein equations)
has been that of modeling a generic single black hole  long periods of time. 
Present single black hole simulations in 3D have not yet been shown to be generically
stable. There are limited instances of stability based on the outer boundary
choice and placement\cite{BlendingStuff,CookScheelstuff}.
Most simulations run just beyond a few hundreds $M$ based on outer
boundary placement and binary black hole simulations run for about $20-50M$ 
before the codes either crash or the entire grid is inside the event horizon. 
In some cases, the reason of the crash is well understood. For instance, the 
use of singularity avoiding slices lead to the presence of steep gradients 
which eventually can no longer be handled by the codes. A solution to this 
problem is to ``excise'' the singularity from the computational 
domain\cite{unruh}. Unfortunately, in most cases, it is not clear what the 
main reasons behind the crash are and consequently addressing the problem 
becomes cumbersome. In attempting to deal with this issue there are several
possible avenues to either remove or provide an understanding of the source
of problems. These avenues can be divided in 
the following way:
\begin{enumerate}
\item{Choice of formulation of Einstein equations;}
\item{Choice of gauge;} 
\item{Numerical implementations.}
\end{enumerate}
Avenue (1) is motivated by the difficulties encountered in achieving long term 
evolutions with the ADM formulation, which historically has been the main tool 
in Numerical Relativity. Several formulations exist in the literature that 
exhibit properties like hyperbolicity\cite{hyperbolic}, the equations are 
expressed in a flux conservative form\cite{fluxconserv} and/or try to separate 
transverse modes\cite{confadm1,baumshap,confadm2}. Avenue (2) is based on the 
fact that, in principle, a coordinate system could be chosen such that the 
fields vary slowly in time; hence, the simulations would be better behaved. 
Conditions to achieve such coordinates have been presented in the 
literature\cite{gauge}. Lastly, avenue (3) highlights the need for a more 
profound understanding of the numerical implementation of the evolution 
equations. Algorithms specifically tailored to deal with the equations
under study could pave the way to better behaved simulations (for instance, 
compare with the implementations that deal with the fluid equations and their 
`historical evolution' from crude implementations in early simulations to high 
resolution shock capturing schemes in present state of the art codes).

Our present work focuses primarily on avenue (1); although avenues (2) and (3)
also play a role since notable improvements are achieved with specific 
gauge choices and the use of causal differencing algorithms. 
We compare results obtained from the use of black hole excision with two 
related forms of the standard ADM 3+1 equations; focusing on the use of
the CADM system with excision techniques in spherical symmetry.

The main motivation behind the comparison with the conformal ADM is the report
by many groups that robust implementations have been achieved in linearized 
gravity, gravitational wave spacetimes, systems containing matter, etc\cite{nakamura,baumshap,confpots_tests}. 
However, so far, it has only been used to model black hole spacetimes using 
singularity avoiding slices\cite{bruegman}. As it is widely accepted, these types of
slicings are useful when the desired simulation time is rather short. In 
order to model black hole spacetimes for long periods of time, singularity 
excision must be employed.  To study the feasibility of excision in this 
formulation and to analyze its advantages and disadvantages with respect to 
the traditional ADM formulation (where excision techniques have been used for 
several years already\cite{huqlett,matt}), we present a $1D$ study and compare 
results obtained with both approaches. We start with a brief review of the 
formulations in section II. In section III, we rewrite the system
of equations in a way convenient for causal differencing and describe how 
this techniques is implemented. In section IV we compare simulations of 
a Schwarzschild black hole and show how the ADM formulation yields longer term 
evolution {\it unless} the trace of the extrinsic curvature is frozen in time, 
in which case CADM yields better behaved evolutions than the ADM formulation, 
we also show how causal differencing indeed gives the expected results in 
terms of stability.  We conclude in section V with a brief 
discussion.


\section{Formulation}
\label{sec:formulation}

The standard ADM equations, in the form most commonly used in numerical
relativity, are\cite{mtw}:
\begin{mathletters}
\label{alleqs}
\begin{eqnarray}
\frac{d}{dt} \; \gamma_{ij} &=& -2\alpha K_{ij}, \label{eqa} \\
\frac{d}{dt} \; K_{ij} &=& -D_{i}D_{j}\alpha + \alpha \left(
\rule{0mm}{4mm} R_{ij} + K K_{ij} \right. \nonumber \\
& & \left.  - 2 K_{ik}K^{k}{}_{j} - {}^{(4)}R_{ij} \right),
\label{eqb}
\end{eqnarray}
\end{mathletters}
\noindent with
\begin{equation}
\frac{d}{dt} = \partial_t - \cal{L}_{\beta} \, ,
\end{equation}
where $\cal{L}_{\beta}$ is the Lie derivative along the
shift vector $\beta^i$; $R_{ij}$ is the Ricci tensor and $D_{i}$
the covariant derivative associated with the three-dimensional metric
$\gamma_{ij}$.  

The conformal ADM equations~\cite{confadm1,baumshap} are obtained from the ADM
ones by (I) making use of a conformal decomposition of the three-metric as
\begin{equation}
\tilde \gamma_{ij} = e^{- 4 \phi} \gamma_{ij} \, \, \, \mbox{with} \, \, \, e^{4 \phi} = \gamma^{1/3} \equiv \det(\gamma_{ij})^{1/3} \, ;
\end{equation}
(hence $det(\tilde \gamma) = 1$); (II) decomposing 
the extrinsic curvature into its trace and trace-free components.
The trace-free part of the extrinsic curvature $K_{ij}$, defined by
\begin{equation}
A_{ij} = K_{ij} - \frac{1}{3} \gamma_{ij} K,
\end{equation}
and $K = \gamma^{ij} K_{ij}$ is the trace of the extrinsic
curvature and (III) further conformally decomposing $A_{ij}$ as:
\begin{equation}
\tilde A_{ij} = e^{- 4 \phi} A_{ij} \, .
\end{equation}
In terms of these variables, Einstein equations in vacuum are~\cite{confadm1,baumshap}
\begin{mathletters}
\label{cadm_alleqs}
\begin{eqnarray}
\frac{d}{dt} \tilde{\gamma}_{ij} &=& - 2 \alpha \tilde{A}_{ij} ,
\label{cadm_alleqsa}
\\
\frac{d}{dt} \phi &=& - \frac{1}{6} \alpha K \, ,
\label{cadm_alleqsb}
\\
\frac{d}{dt} K &=& - \gamma^{ij} D_i D_j \alpha  + \alpha \left[
\tilde{A}_{ij} \tilde{A}^{ij} + \frac{1}{3} K^2 \right] ,
\label{cadm_alleqsc}
\\
\frac{d}{dt} \tilde{A}_{ij} &=& e^{-4 \phi} \left[
 - D_i D_j \alpha + \alpha R_{ij} \right]^{TF}
\noindent \nonumber \\
&& + \alpha \left( K \tilde{A}_{ij} - 2 \tilde{A}_{il} \tilde{A}_j^l
\right) ,
\label{cadm_alleqsd}
\end{eqnarray}
\end{mathletters}
where the Hamiltonian constraint was used to eliminate the Ricci scalar in equation (\ref{cadm_alleqsc}).
Note that with the conformal decomposition of the three--metric, the Ricci
tensor now has two pieces, which are written as
\begin{equation}
R_{ij} = \tilde{R}_{ij} + R^{\phi}_{ij} .
\end{equation}
The ``conformal-factor'' part $R^{\phi}_{ij}$ is given directly by
straightforward computation of derivatives of $\phi$:
\begin{eqnarray}
R^{\phi}_{ij} &=& - 2 \tilde{D}_i \tilde{D}_j \phi - 2 \tilde{\gamma}_{ij}
\tilde{D}^l \tilde{D}_l \phi \noindent \\
&& + 4 \tilde{D}_i \phi \; \tilde{D}_j \phi - 4 \tilde{\gamma}_{ij}
\tilde{D}^l \phi \; \tilde{D}_l \phi ,
\end{eqnarray}
while the ``conformal'' part $\tilde{R}_{ij}$ can be computed in the
standard way from the conformal three--metric $\tilde \gamma_{ij}$.

To this point, the equations have been written by a trivial algebraic
manipulation of the ADM equations
in terms of the new variables. The non-trivial part comes into play by introducing
what Ref.~\cite{baumshap} calls the ``conformal connection functions'':
\begin{equation}
\tilde{\Gamma}^i := \tilde{\gamma}^{jk} \tilde{\Gamma}^i_{jk} =
 - \tilde{\gamma}^{ij}_{~~,j} ,
\end{equation}
where the last equality holds since the determinant of the conformal
three--metric $\tilde \gamma$ is unity.
Using the conformal connection functions, the Ricci tensor is
written as:
\begin{eqnarray}
\tilde R_{ij} & = & - \frac{1}{2} \tilde \gamma^{lm}
        \tilde \gamma_{ij,lm}
        + \tilde \gamma_{k(i} \partial_{j)} \tilde \Gamma^k
        + \tilde \Gamma^k \tilde \partial_{(j} \tilde \gamma_{i)k} \nonumber \\
& &  - \tilde \gamma^{kl}_{(,j} \tilde \gamma_{i)l,k} - \tilde \Gamma^l \tilde \Gamma_{ijl}
     - \Gamma^l_{kj} \Gamma^{k}_{li} \, .
\end{eqnarray}
Where $\tilde \Gamma^i$ are to be considered independent variables whose
evolution equations are obtained by a simple commutation of derivatives.
\begin{eqnarray}
\frac{\partial}{\partial t} \tilde \Gamma^i
&=& - \frac{\partial}{\partial x^j} \Big( 2 \alpha \tilde A^{ij}
- 2 \tilde \gamma^{m(j} \beta^{i)}_{~,m} \noindent \nonumber \\
&& + \frac{2}{3} \tilde \gamma^{ij} \beta^l_{~,l}
        + \beta^l \tilde \gamma^{ij}_{~~,l} \Big) .
\label{eq:evolGamma}
\end{eqnarray}

As proposed in Ref.~\cite{baumshap} the
divergence of $\tilde A^{ij}$ is replaced with the help of the momentum
constraint to obtain:
\begin{eqnarray}
\frac{\partial}{\partial t} \tilde \Gamma^i
& = & - 2 \tilde A^{ij} \alpha_{,j} + 2 \alpha \Big(
        \tilde \Gamma^i_{jk} \tilde A^{kj} -
        \frac{2}{3} \tilde \gamma^{ij} K_{,j}
        + 6 \tilde A^{ij} \phi_{,j} \Big)
\nonumber \\
& & +
\beta^l \tilde \Gamma^i_{,l} + \frac{1}{3} \tilde \gamma^{mi} \beta^j_{,mj}
+ \tilde \gamma^{mj} \beta^i_{,mj} - \tilde \Gamma^m \beta^i_{,m}
+\frac{2}{3} \tilde \Gamma^i \beta^l_{,l} .
\label{eq:evolGamma2}
\end{eqnarray}

With this reformulation, in addition to the evolution equations for
the conformal three--metric $\tilde \gamma_{ij}$~(\ref{cadm_alleqsa}) and
the conformal-traceless extrinsic curvature variables $\tilde
A_{ij}$~(\ref{cadm_alleqsd}), there are evolution equations for the
conformal factor $\phi$~(\ref{cadm_alleqsb}), the trace of the
extrinsic curvature $K$~(\ref{cadm_alleqsc}) and the conformal connection
functions $\tilde \Gamma^i$~(\ref{eq:evolGamma2}).

\section{Causal differencing implementation}
Causal differencing, as explained in\cite{seidelsuen,alcubschutz,gundwalk,mark2,bbhmanual}, provides a
straightforward way to integrate the evolution equations while
preserving (and taking advantage of) the causal structure of the spacetime
under consideration. In the approach used in the present work
we follow the strategy described in\cite{bbhmanual}. First, the Lie derivative along $\beta$ 
is split and  terms containing derivatives of $\beta$ are moved 
to the right hand side. Then, the ADM system of equations is reexpressed
as
\begin{mathletters}
\label{all_adm}
\begin{eqnarray}
\partial_o {\gamma}_{ij} &=& - 2 \alpha {\gamma}_{ij} + 2 {\gamma}_{l(i} \beta^l_{,j)} \label{all_adma} \, , \\
\partial_o K_{ij} &=&  D_{i}D_{j}\alpha + \alpha \left(
\rule{0mm}{4mm} R_{ij} + K K_{ij} \right. \nonumber \\
& & \left.  - 2 K_{ik}K^{k}{}_{j} - {}^{(4)}R_{ij} \right) + 2 K_{l(i} \beta^l_{,j)} \label{all_admb}  \, ;
\end{eqnarray}
\end{mathletters}
and the CADM system of equations then reduces to
\begin{mathletters}
\label{all_cadm}
\begin{eqnarray}
\partial_o \tilde{\gamma}_{ij} &=& - 2 \alpha \tilde{A}_{ij} 
+ 2 \tilde{\gamma}_{l(i} \beta^l_{,j)} \, ,
\label{all_cadma}
\\
\partial_o \phi &=& - \frac{1}{6} \alpha K + \frac{1}{6} \beta^i_{,i} \, , 
\label{all_cadmb}
\\
\partial_o  K &=& - \gamma^{ij} D_i D_j \alpha  + \alpha \left[
\tilde{A}_{ij} \tilde{A}^{ij} + \frac{1}{3} K^2 \right] ,
\label{all_cadmc}
\\
\partial_o \tilde{A}_{ij} &=& e^{-4 \phi} \left[
 - D_i D_j \alpha + \alpha R_{ij} \right]^{TF}
\noindent \nonumber  \\
&& + \alpha \left( K \tilde{A}_{ij} - 2 \tilde{A}_{il} \tilde{A}_j^l
\right)  \nonumber \\
&& + 2 \tilde{A}_{k(j} \beta^k_{,i)} - \frac{2}{3} \tilde{A}_{ij} \beta^k_{,k} \, ,
\label{all_cadmd}
\\
\partial_o \tilde \Gamma^i
& = & - 2 \tilde A^{ij} \alpha_{,j} + 2 \alpha \Big(
        \tilde \Gamma^i_{jk} \tilde A^{kj} -
        \frac{2}{3} \tilde \gamma^{ij} K_{,j}
        + 6 \tilde A^{ij} \phi_{,j} \Big)
\nonumber \\
& & +
 \frac{1}{3} \tilde \gamma^{mi} \beta^j_{,mj}
+ \tilde \gamma^{mj} \beta^i_{,mj} - \tilde \Gamma^m \beta^i_{,m}
+\frac{2}{3} \tilde \Gamma^i \beta^l_{,l} \,  .
\label{all_cadme}
\end{eqnarray}
\end{mathletters}
where $\partial_o \equiv \partial_t - \beta^i \partial_i$.

\begin{figure}
\centerline{\epsfxsize=200pt\epsfbox{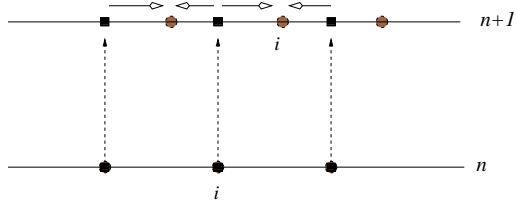}}
\caption{
Illustration of the causal differencing strategy. First the integration
proceeds along the dashed lines to obtain values in the $n+1$ level (at
filled square points).
Then, an interpolation is carried out to obtain values on the grid
points (filled circles).
In the graph, as an example, a second order interpolation
(indicated with arrows) provides values on the i-th grid point. }
\label{causalsteps}
\end{figure}

Finally the numerical implementation of the equations is split into two steps. 
First, the equations are evolved along the normal to the hypersurface (at constant $t$) $n^a=\partial_t^a
-\beta^i \partial_i^a$. In the second step, an interpolation is carried over to obtain
values on grid coordinate locations (see Fig \ref{causalsteps}). Note that the
two systems of equations have in this form the same basic structure; hence, simple modifications
to an ADM code with excision (like AGAVE\cite{agave}) will enable the use of already developed excision
modules with the CADM equations in a straightforward manner\footnote{These modifications
are in place in the AGAVE code and currently being tested in 3D}.

\subsection{Numerical Implementation}
In the numerical implementation of the CADM it is convenient to 
introduce an intermediate variable $F$ such that 
$\phi \equiv 1/4 \ln(F)$ and evolve $F$ instead of $\phi$. This choice avoids
unnecessary handling of exponential and logarithmic functions 
thus preventing loss of accuracy; hence, the equation for $F$ is
\begin{equation}
\partial_o F = - \frac{2}{3} \alpha  F K + \frac{2}{3} F \beta^i_{,i}
\end{equation}
A second order finite difference code has been written to implement both the 
ADM and CADM formulations. The same causal differencing algorithms have been
applied to both formulations. These algorithms involve at their core, an
interpolation for every grid point in the computational domain. Near an
excision boundary the choice of interpolation order and stencil choice
becomes important. We allow a choice of second, third and fourth order
interpolations in order to study possible practical approaches  \footnote{This 
code is publicly available and can be requested from the authors.}.

\section{Applications}
To compare evolutions with the above formulations we pick as a particular
example the Schwarzschild spacetime (and linear perturbations of it). In order
to implement excision, a slicing must be chosen such that surfaces of constant
time ``penetrate'' the horizon. The ingoing Eddington Finkelstein coordinates
define hypersurfaces satisfying this condition; in terms of them, the line
element reads
\begin{equation}
ds^2 = -\left(1-\frac{2M}{r}\right) dt^2 + \frac{4M}{r} dt\, dr + \left(1+ \frac{2M}{r}\right) dr^2 + r^2 d \Omega^2\,,
\label{ief}
\end{equation}
where $d\Omega^2 = d\theta^2+\sin^2\theta\,d\phi^2$. The lapse and shift vector are therefore:
\begin{eqnarray}
\alpha = \sqrt{\frac{r}{r+2M}} \, , \, \beta^i = \frac{2M}{r+2M} \delta^i_r \, ,
\end{eqnarray}
The basic ADM variables read
\begin{eqnarray}
\gamma_{rr} &=& 1 + 2 \frac{M}{r} \, ; \,
\gamma_{\theta \theta} = r^2 = \frac{\gamma_{\phi \phi}}{\sin^2\theta} \nonumber\\
K_{rr} &=& -\frac{2M}{r^3} (r+M) \alpha \, ; \,
K_{\theta \theta} = 2 M \alpha = \frac{K_{\phi \phi}}{\sin^2\theta} \nonumber \,.
\end{eqnarray}
and the CADM variables
\begin{eqnarray}
\tilde \phi &=& \frac{1}{4} \ln(\left[ (r+2M) r^3 \sin^2 \theta \right]^{1/3}) \, ; \,
K = \frac{2M \alpha }{r^2}\frac{(r+3M)}{(r+2M)} \nonumber \\
\tilde \gamma_{rr} &=& \frac{r+2M}{r^2 \left[ (r+2M) \sin^2 \theta \right]^{1/3}} \, ; \,
\tilde \gamma_{\theta \theta} = \frac{r}{\left[ (r+2M) \sin^2 \theta \right]^{1/3}} 
= \frac{\tilde \gamma_{\phi \phi}}{\sin^2\theta} \nonumber\\
\tilde A_{rr} &=& -\frac{4 M}{3} \frac{\alpha (2r+3M)}{r^4 \left[(r+2M) \sin^2 \theta \right]^{1/3}} \nonumber \\
\tilde A_{\theta \theta} &=& \frac{2M}{3} \frac{\alpha (2r+3M)}{r (r+2M) \left[ (r+2M)\sin^2 \theta \right]^{1/3}}
 = \frac{\tilde A_{\phi \phi}}{\sin^2\theta} \nonumber \, \\
\tilde \Gamma^r &=& -\frac{4}{3} \frac{r^3 (r+3M) \sin^2 \theta }{r^2 (r+2M)^{5/3}} \nonumber \\
\tilde \Gamma^{\theta} &=& -\frac{2}{3} \frac{r (r+2M) \cos \theta  }{r^2 (r+2M)^{2/3} \sin^{2/3} \theta }
\nonumber \\
\tilde \Gamma^{\phi} &=& 0 \nonumber \, .
\end{eqnarray}
Note that some of the quantities are functions of $\theta$. In our spherically
symmetric implementation of these equations we have explicitly expressed each variable
as a function of $r$ times the exact function of the angle $\theta$. For instance we write
\begin{equation}
\tilde \gamma_{\theta \theta}= h_{\theta \theta}(r)/\sin^2\theta \, .
\end{equation}
Proceeding this way allows for the explicit appearance of $\theta$ to drop out of the 
equations, providing at the end of the day, a truly $1D$ system of
equations corresponding to spherical symmetry.

\subsection{Comparison}
Extended tests were performed with both codes (under the same conditions) to 
understand the robustness of each formulation with 
excision. As has been observed in previous work\cite{baumshap,confpots_tests}, 
CADM gives long term evolutions when the evolution of $K$ is ``frozen"; ie the 
equation for $K$ is not evolved or the value of $K$ is fixed by the choice of
a slicing that leaves $K$ fixed (for instance, maximal slicing that
fixes $K=0$.). On the other hand, longer term evolutions have also
been achieved
with an area locking gauge in the ADM formulation\cite{texas2000}.
We then perform three basic tests:
\begin{itemize}
\item{{\it Fully free evolution:} All equations corresponding to each system 
are integrated without imposing any further condition}
\item{{\it``Locked" evolution:} Conditions on some of the field variables are enforced
(see below).}
\item{{\it``Perturbed" evolution:} Same as the ``locked'' case but considering linear
perturbations of Schwarzschild spacetime as initial data}
\end{itemize}
In all these tests, we study the dependence of the obtained
solution under discretization size and location of the outer boundary.
The inner boundary is placed at $r=M$
and the outer boundary is varied (placed at $r=nM$) while keeping $\Delta r=const$.
Outer boundary data are provided by `blending'\cite{roberto} the
numerical solution to the analytical one. This choice reduces
gradients and second derivatives at the boundary allowing for a clean
evolution without much reflections from the outer boundary.

\subsubsection{Fully free evolution}
In this case, all equations corresponding to systems (\ref{all_adm},\ref{all_cadm})
 are evolved
and the obtained solutions are compared. We use the Hamiltonian and
momentum constraint as monitors of the quality of the evolution. Our results 
can be summarized as follows. For the ADM formulation we observed that stable
evolutions are obtained if $n<6$ while for larger $n>6$ the solution exhibits
exponentially growing modes. It is worth emphasizing
that the evolution is not unstable under the strict sense (i.e. the solution
can be bounded from above by an exponential\cite{kreiss}.). However, the presence of
this exponential mode will likely spoil any long term simulation.
For the CADM system, irrespective of the value of $n$ exponential modes are
clearly present in the solutions. 

\begin{figure}
\centerline{\epsfxsize=400pt\epsfbox{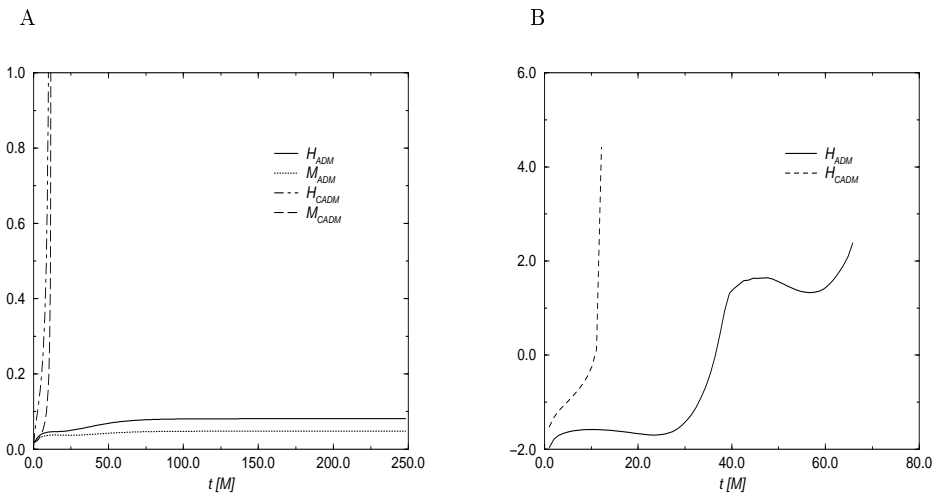}}
\caption{
The $L_2$ norm of the Hamiltonian and momentum constraints
for both formulations (discretization size $\Delta r = M/10$)
for the case where the domain of integration
is $[M,4M]$ (A) and $[M,9M]$ (B). In A, the evolution obtained with 
the ADM formulation does not show the presence of an exponentially
growing mode like the one obtained with the CADM approach. However,
for the larger domain, (B), solutions obtained with both formulations
are exponentially growing.}
\label{fig_testa}
\end{figure}

These results are illustrated in Fig.\ref{fig_testa}, which shows the $L_2$ norm
of the Hamiltonian and momentum constraints of the solutions obtained with
both formulations when $n=4$. Clearly, the solution obtained with the ADM
is better than that obtained with the CADM. Figure \ref{fig_testa}, also displays
the $L_2$ norm of the Hamiltonian constraint for the case $n=9$, although the solution
obtained with the ADM formulation can be considered better than that from the CADM, both
grow exponentially.

\subsubsection{Locked evolution}
\begin{figure}
\centerline{\epsfxsize=400pt\epsfbox{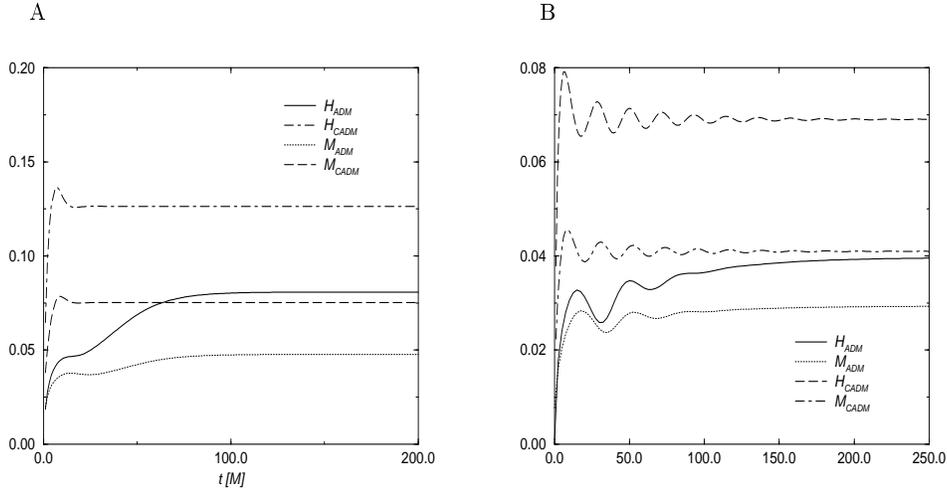}}
\caption{
The $L_2$ norm of the Hamiltonian and momentum constraints
for both formulations (with $\Delta r= M/10$) for the case where the domain 
of integration is $[M,4M]$ (A). Neither formulation displays
exponential modes in this domain and the ADM one yields more
accurate results. In B, the $L_2$ norm of the Hamiltonian
and momentum constraints of the solution obtained with the
CADM  and an ``area-locked'' ADM evolution
(in the $[M,9M]$ domain) is shown, a transient oscillatory
behavior is present at earlier stages and the solutions then settle to a constant value.}
\label{fig_testb}
\end{figure}

It has been observed in the literature\cite{confpots_tests} that in the case
where $K$ is fixed in time very long term 
evolutions can be obtained with the CADM system. It is interesting to see
what this condition implies, note that if $K$ does not vary, then, $\phi$ will
remain unchanged and therefore the determinant of the three metric $\gamma_{ij}$
will remain independent of time. This, could be regarded as an evolution
that ``locks" the volume and bears some similarity with the so-called ``area-locking"
gauge\cite{seidelsuen,texas2000}.
It is also worth pointing out that a similar strategy can be implemented
in the ADM case as it has been shown in\cite{texas2000}. In this work, by choosing
a gauge that ``locks" the evolution of $g_{\theta \theta}$ the exponentially growing
modes displayed by solutions in domains with $6<n<11$ are removed. 
Figure \ref{fig_testb} shows
the $L_2$ norm of the Hamiltonian and momentum constraints corresponding
to solutions obtained with both formulations for
the choice $n=4$ and $n=9$. In both cases the simulations can be performed
for unlimited time without observing exponential modes. Again, the solution
obtained with the ADM formulation is slightly more accurate than that
provided by the CADM formulation.

It is important to observe that for the CADM case with $K$ frozen, 
evolutions without exponentially
growing modes are obtained with $n$ as large as $16$. For
$n>16$ long term evolutions ($>100M$) display at late
times a clear exponentially growing mode.

\subsubsection{Perturbed evolution}
\begin{figure}
\centerline{\epsfxsize=250pt\epsfbox{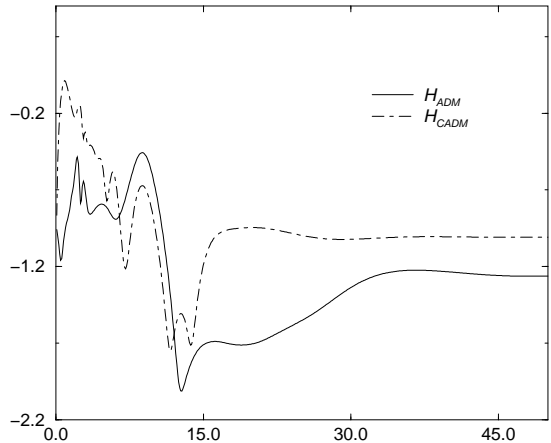}}
\caption{
The $L_2$ norm of the Hamiltonian constraint for the perturbed
evolutions (where $K$ has been frozen in the CADM evolution
while all fields are evolved in the ADM one). After some
initial transient behavior both
settle into an stationary solution.}
\label{pulse}
\end{figure}

In this case, we test the evolutions under perturbations (using a locked
evolution in the CADM case but not in the ADM one). The initial
data corresponds to the analytic value of $\gamma_{rr}$ (or $\tilde \gamma_{rr}$) plus
some
arbitrary pulse of compact support. Of course, this data is unphysical
but we use it to probe for stability of the implementations in a non-trivial
scenario. The amplitude of the pulse is chosen such that it
can be considered a linear perturbation of a Schwarzschild spacetime.
The results obtained with both codes agree with those of the previous
section. Figure \ref{pulse} corresponds to the evolution of
a pulse with compact support in $[3M,5M]$ being evolved in a computational
domain of $[M,6M]$. The $L_2$ norm of the Hamiltonian constraint, after some
initial transient behavior, settles into a stationary regime.

\subsection{Locking evolutions}
\begin{figure}
\centerline{\epsfxsize=250pt\epsfbox{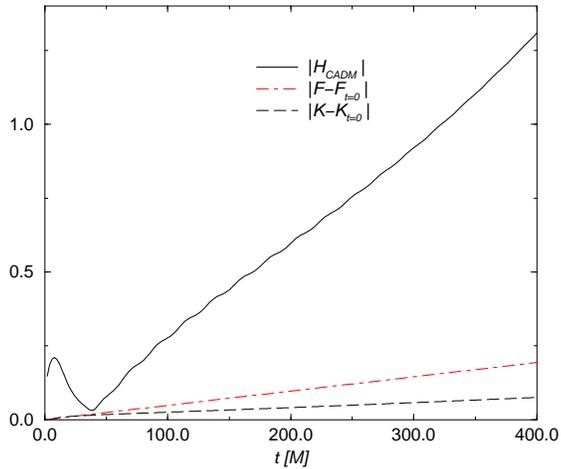}}
\caption{
The $L_2$ norms of the Hamiltonian constraint and the differences
$F-F_t=0$ and $K-K_{t=0}$ vs. time. The evolution proceeds without displaying
exponential modes and the value of $|K-K_{t=0}|$ stays close to zero.}
\label{lockshift}
\end{figure}

As mentioned, in previous work\cite{confpots_tests}, by not evolving the
equation for $K$
very long term evolutions have been achieved with the CADM formulation.
However, choosing to do so is unphysical in generic situations. One would
like to have a prescription where a similar condition can be enforced
without having to not evolve one or more equations. Here, one can use the
gauge freedom of the theory to demand that $\partial_t K = 0$. This in
turns implies a condition of the shift vector from equation (\ref{all_cadmc}), 
\begin{equation}
\beta^i \partial_i K = \gamma^{ij} D_i D_j \alpha  - \alpha \left[
\tilde{A}_{ij} \tilde{A}^{ij} + \frac{1}{3} K^2 \right] \, \label{lockK}.
\end{equation}
This condition is straightforward to implement in 1D but is certainly
more complicated in 3D. Additionally, there is a great deal of ambiguity as
it is only one equation for three variables $\beta^i$. Therefore
two supplementary conditions must be chosen so that (\ref{lockK})
can be used to `freeze' the evolution of $K$. In our present implementaion
we have simply chosen $\beta^A=0$ (with $A=(\theta,\phi)$) and
obtain $\beta^r$ with (\ref{lockK}) as
\begin{equation}
\beta^r =\frac{1}{\partial_r K} \left( \gamma^{ij} D_i D_j \alpha  - \alpha \left[
\tilde{A}_{ij} \tilde{A}^{ij} + \frac{1}{3} K^2 \right] \right) \, \label{lockK_2}.
\end{equation}
A simple way to obtain $\beta^r$ is by
a first order approximation of the rhs of Eq. \ref{lockK_2} (ie. evaluating each term at
the old level). By using this condition, instead of choosing not to evolve $K$,
we were able to obtain evolutions not displaying exponential modes for times 
larger than $250M$ (with resolutions of $\Delta r=M/10$
and finer). Figure \ref{lockshift} illustrates what is obtained in a simulation
with computational domain defined by $[M,11M]$  (with $\Delta r=M/10$). The values
of the $L_2$ norms of the Hamiltonian, the function $F-F_{t=0}$ and value of
$K-K_{t=0}$
are shown as a function of time. Since $\beta^r$ is obtained only as
a first order approximation, $K$ and $F$ are expected to vary during the evolution.
As can be seen in Fig. \ref{lockshift}, both grow linearly but stay fairly close to
zero and the evolution proceeds without displaying an exponential growth.

\subsection{Causal differencing and domain of dependencies}
\begin{figure}
\centerline{\epsfxsize=300pt\epsfbox{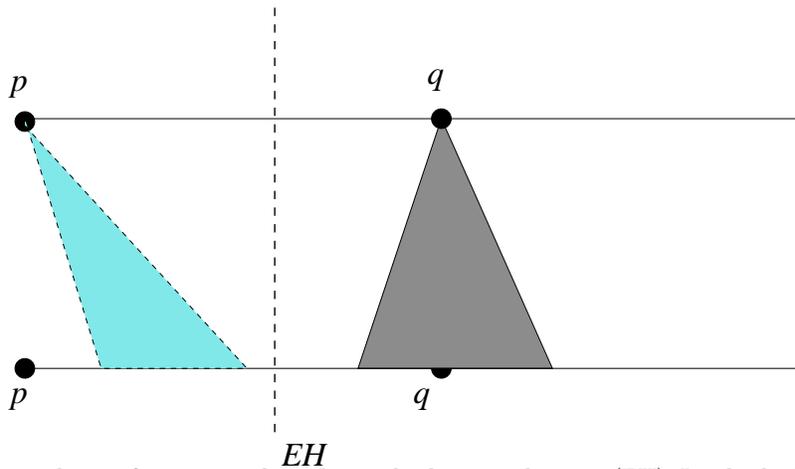}}
\caption{
Domains of dependence of points inside and outside the event horizon (EH).
Inside the EH, the past null cone of $p$ is tilted, therefore the evolution
algorithm does not need the value of the fields at $p$ on the old level.}
\label{pablo}
\end{figure}

\begin{figure}
\centerline{\epsfxsize=250pt\epsfbox{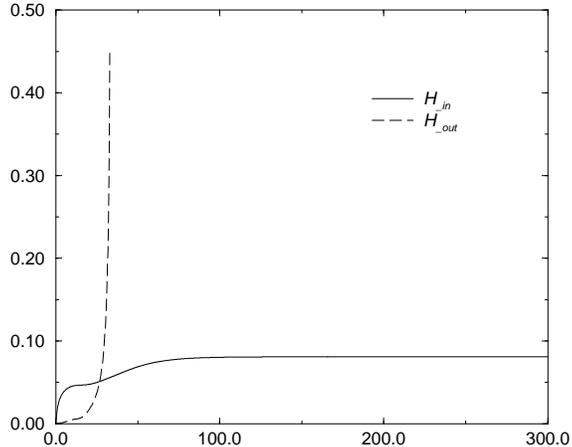}}
\caption{
The $L_2$ norm of the Hamiltonian constraint for the cases where the
inner boundary is inside (at $r=1.5M$ denoted with a solid line) and
outside (at $r=2.2M$ denoted with a dashed line)
the event horizon (using the ADM formulation). Since the latter
does not respect the CFL condition,
the obtained solution is unstable. (The values shown correspond to a 
discretization of size $\Delta r=M/10$ and no qualitative difference
is observed with finer resolutions.)}
\label{pabloplace}
\end{figure}

As a last point, it is interesting to see how causal differencing
is indeed providing a correct way to discretize the equations taking
advantage of the causal properties of the spacetime. The fact that
the null cones (or the causal domain of dependence) are tilted inside
the horizon, allows for a stable numerical implementation where inner boundary
data need not be provided if the inner boundary is inside the black hole (see
Fig. \ref{pablo}).
This is possible because the numerical domain of dependence naturally
contains the domain of dependence of the inner boundary
point. This is of course, not true if the innermost point is outside the
event horizon. In order to illustrate this fact we compare $2$ cases
(with both formulations) where the innermost point is placed inside
or just outside the event horizon.  As illustrated in Fig. \ref{pabloplace}, 
while the solution obtained with the inner boundary
inside the black hole is stable, the other, as expected is unstable.

\section{Conclusions}
The results presented in this work show that excision techniques
can be straightforwardly used in the CADM formulation directly
from the structures developed for the ADM formulation.

The ADM formulation is superior to the CADM both in accuracy and
total time evolution when the evolution of $K$ is not locked
in CADM. When locking is implemented, then CADM is better
than ADM as the solution obtained with the CADM formulation
does not display exponential modes with the outer boundary placed
as far as $16M$. On the other hand, evolutions with the ADM formulation
display exponential modes with the outer boundary placed at $11M$ and
beyond. Additionally, for the case where outer
boundaries are placed `very' far, although exponentially
growing solutions are present in solutions obtained
with both formulations, ADM simulations crash at earlier
times than CADM. 

It is worth remarking again that in both formulations, implementing a gauge
that minimizes the changes in some of the fields (like $g_{\theta \theta}$
in the ADM formulation or $K$ in the CADM one) dramatically improves the
evolutions in 1D. In the 3D case, the use of `area or circunference locking'\footnote{ 
ie. controling the determinant of the angular part of the metric (area locking)
or  $g_{\theta \theta}$ (circunference locking).} is indeed more
complicated than locking $K$, simply by the fact that in the former case one
is trying to control a tensor component while in the latter a scalar. Thus,
locking $K$ is likely to have an easier and probably more general
implementation than area locking (although in cases where the final black
hole is close to a non spinning one, this implementation is rather
straightforward). Controling the evolution of $K$ demans a condition like 
that given by Eq (\ref{lockK}), and two extra conditions on $\beta^i$ will be 
required. An option that could mimic the 1D implementation would be to
foliate the 3D hypersurfaces with a sequence of 2-surfaces defined by
$\Theta=const$ (with $\Theta$ the expansion of outgoing null rays). Once this
foliation is obtained, the shift vector $\beta^i$ could be decomposed as 
\begin{equation}
\beta^i = \beta^i_{|\ } + \beta^i_{\bot}
\end{equation} 
with $\beta^i_{|\ }$ ($\beta^i_{\bot}$) parallel (perpendicular) to the
normal of the 2-surfaces. Thus, the two further conditions can be chosen
such that $\beta^i_{\bot}=const$.
This will thus minimize changes in transversal directions
and will resemble that we have used in the 1D case. Of course, this is
just one possible approach and further studies will be required to 
obtain a $K$ fixing condition that leads to a practical implementation.

In conclusion,  implementing singularity excision techniques in the
CADM formulation is straightforward. The usefulness of this implementation
depends on implementing a gauge controlling
the behavior of $K$. Assuming this can be achieved, 
CADM appears to be capable of providing more robust
simulations than ADM when the outer boundary is placed farther than $11M$
from the final black hole of mass $M$, if the outer boundary is closer,
then the ADM formulation provides evolutions as stable as the CADM one
but with better accuracy. 

As a last remark, it is important to stress that we have only implemented 
the causal differencing algorithm described in\cite{bbhmanual,mark2} since
at present is the only one fully implemented in $3D$. Other alternatives
have been proposed\cite{alcubschutz,gundwalk,movebh}; due to the restriction
to spherical symmetry it is likely that the application of these will yield
similar results to those presented in this work.

\acknowledgments
This work was supported by NSF PHY 9800725 to the University of Texas at Austin
and NSF 9800970 and NSF 9800973 to the Pennsylvania State University.
We thank  D.~Neilsen, P. Marronetti, R. Matzner, P. Laguna and J.~Pullin for
helpful comments and suggestions. D.G. is an Alfred P. Sloan Scholar




\begin{references}
\bibitem{BlendingStuff}
M. Huq. Proceedings of the BBH Grand Challenge Meeting.
Pittsburgh, PA 1997.

\bibitem{CookScheelstuff}
G. Gook and M. Shceel. Proceedings of the BBH Grand Challenge Meeting.
Austin, TX 1998.

\bibitem{unruh}
J. Thornburg, {\it Class. and Quantum Grav.}, {\bf 14}, 1119 (1987).

\bibitem{hyperbolic}
For instance, H. Friedrich, Proc. Roy. Soc. Lond. {\bf A375},  169  (1981). 
S. Frittelli, O. Reula, {\it Phys. Rev. Lett}, {\bf 76}, 4667, (1996).
A. Anderson and J. York, {\it Phys. Rev. Lett}, {\bf 82}, 4384, (1999).
(For a detailed list of hyperbolic formulations see O. Reula,
Living Reviews in Relativity (1998)).

\bibitem{fluxconserv}
C. Bona, J. Masso, E. Seidel and J. Stela,
{\it Phys. Rev. D.}, {\bf 56}, 3405 (1997). 

\bibitem{confadm1}
M. Shibata and T. Nakamura {\it Phys. Rev. D.}, {\bf 52}, 5428 (1995).

\bibitem{baumshap}
T. Baumgarte and S. Shapiro. {\it Phys. Rev. D.}, {\bf 59}, 024007 (1999). 

\bibitem{confadm2}
S. Frittelli and O. Reula, gr-qc/9904048 (1999).

\bibitem{gauge}
L.L.Smarr and J.W.York, {\it Phys. Rev. D.}, {\bf 17}(10), 2529-2551 (1978).
P.R.~Brady, J.D.~Creighton and K.S.~Thorne, {\it Phys. Rev. D.}, {\bf 58}, 061501 (1998).
C. Gundlach and D. Garfinkle., {\it Class. Quant. Grav.}, {\bf 16} (1999).

\bibitem{nakamura}
K. Oohara and T. Nakamura `3D General Relativistic Simulations
of Coalescing Binary Neutron Stars', astro-ph/9912085 (1999).

\bibitem{confpots_tests}
M. Alcubierre, G. Allen, B. Br\"{u}gmann, T. Dramlitsch,
J. Font, P. Papadopoulos, E. Seidel, N. Stergioulas,
W. M. Suen and R. Takahashi. `Towards a stable
numerical evolution of strongly gravitating systems
in General Relativity'. gr-qc/0003071, (2000).

\bibitem{bruegman}
B. Br\"{u}gmann, {\it Int. J. Mod. Phys.} {\bf D8}, 85  (1999).

\bibitem{huqlett}
The Binary Black Hole Grand Challenge Alliance, 
{Phys. Rev. Letters}, {\bf 80}, 2512 (1998). 

\bibitem{matt}
R. Marsa and M. Choptuik, {\it Phys. Rev. D.}, {\bf 54}, 4929 (1996).


\bibitem{mtw}
C. Misner, K.~S. Thorne, and J. Wheeler, {\em Gravitation} (W. H. Freeman and
Co., San Francisco, 1973).
 
\bibitem{seidelsuen} 
E. Seidel, W.M. Suen,  {\it Phys. Rev. Letters} {\bf 69} (1992).

\bibitem{alcubschutz}
M. Alcubierre and B. Schutz, {\it J. Comp. Phys.} {\bf 112} (1994).
 
\bibitem{gundwalk}
C. Gundlach and P. Walker, {Class Quant Grav} {\bf 16} 991-1010 (1999) 
 
\bibitem{mark2}
M. Scheel, T. Baumgarte, G. Cook, S. Shapiro and S. Teukolsky,
{\it Phys. Rev. D.}, {\bf 56}, 6320 (1997).
 
\bibitem{bbhmanual}
 M. Huq and R. Matzner, ``BBHGC Alliance Cauchy Code Documentation", 
{\it unpublished}. 

\bibitem{agave} 
http://www.astro.psu.edu/users/nr/Agave/.

\bibitem{texas2000}
L. Lehner, M. Huq, M. Anderson, E. Bonning, D. Schaefer and R. Matzner,
to appear in {\it Phys. Rev. D.}, (2000).

\bibitem{roberto}
R. Gomez. Proceedings of the BBH Grand Challenge Meeting.
Los Alamos, NM 1997.

\bibitem{kreiss}
B. Gustaffson, H. Kreiss and J. Oliger. {\em Time Dependent Problems
and Difference Methods} (John Wiley \& Sons, New York, 1995). 

\bibitem{movebh}
R. Gomez, L. Lehner, R. Marsa and J. Winicour, 
{\it Phys. Rev. D.}, {\bf 57}, 4778 (1998).

\end{references}

\end{document}